\documentclass[aps,prb,twocolumn,amsmath,amssymb,superscriptaddress,longbibliography]{revtex4-2}
\usepackage{graphicx} 
\newcommand{\be}{\begin{equation}}
\newcommand{\ee}{\end{equation}}
\newcommand{\bg}{\begin{equation}}
\newcommand{\eg}{\end{equation}}
\newcommand{\bdm}{\begin{displaymath}}
\newcommand{\edm}{\end{displaymath}}
\newcommand{\bea}{\begin{eqnarray}}
\newcommand{\eea}{\end{eqnarray}}
\newcommand{\beas}{\begin{eqnarray*}}
\newcommand{\eeas}{\end{eqnarray*}}
\newcommand{\ba}{\begin{array}}
\newcommand{\ea}{\end{array}}
\newcommand{\nn}{\nonumber}

\newcommand{\bfg}{\begin{figure}}
\newcommand{\efg}{\end{figure}}

\newcommand{\tl}{\tilde}

\newcommand{\fr}{\frac}

\newcommand{\mb}{\mbox}

\newtheorem{lm}{Lemma}
\newtheorem{cl}{Corollary}
\newtheorem{df}{Definition}

\newcommand{\blm}{\begin{lm}}
\newcommand{\elm}{\end{lm}}
\newcommand{\bcl}{\begin{cl}}
\newcommand{\ecl}{\end{cl}}
\newcommand{\bdf}{\begin{df}}
\newcommand{\edf}{\end{df}}
\newcommand{\brk}{\begin{rm}}
\newcommand{\erk}{\end{rm}}

\newcommand{\vst}{\vspace*}

\newcommand{\DD}{{\cal D}}

\newcommand{\om}{\omega}
\newcommand{\Om}{\Omega}

\newcommand{\bt}{\beta}
\newcommand{\dt}{\delta}

\newcommand{\veps}{\varepsilon}
\newcommand{\vphi}{\varphi}
\newcommand{\ld}{\lambda}

\newcommand{\gm}{\gamma}

\newcommand{\Gm}{\Gamma}

\newcommand{\vnab}{\mbox{\boldmath $\nabla$}}

\newcommand{\vA}{{\bf A}}

\newcommand{\vB}{{\bf B}}

\newcommand{\vC}{{\bf C}}

\newcommand{\vE}{{\bf E}}
\newcommand{\ve}{{\bf e}}

\newcommand{\vH}{{\bf H}}

\newcommand{\vm}{{\bf m}}

\newcommand{\vn}{{\bf n}}

\newcommand{\vP}{{\bf P}}
\newcommand{\vp}{{\bf p}}
\newcommand{\vQ}{{\bf Q}}

\newcommand{\vvr}{{\bf r}} 
\newcommand{\vS}{{\bf S}}

\newcommand{\vY}{{\bf Y}}

\begin{document}

\title{Generation of nearly pure and highly directional magnetic light in fluorescence of rare earth ions}

\author{Anton D. Utyushev}
 \affiliation{School of Physics and Engineering, ITMO University, 197101, Saint-Petersburg, Russia}
 \email{ad.utyushev@gmail.com}
\author{Roman Gaponenko}
 \affiliation{School of Physics and Engineering, ITMO University, 197101, Saint-Petersburg, Russia}
\author{Song Sun}
 \affiliation{Microsystem and Terahertz Research Center, China Academy of Engineering Physics, No. 596, Yinhe Road, Shuangliu, Chengdu, 610200, China}
 \affiliation{Institute of Electronic Engineering, China Academy of Engineering Physics, Mianyang, 621999, China}
\author{Alexey A. Shcherbakov}
 \affiliation{School of Physics and Engineering, ITMO University, 197101, Saint-Petersburg, Russia}
\author{Alexander Moroz}
 \affiliation{Wave-scattering.com}
\author{Ilia L. Rasskazov}
 \affiliation{SunDensity Inc., Rochester, NY 14604, USA}

\date{\today}

\begin{abstract}
A thorough analysis of the emission via the magnetic dipole (MD) transition, called magnetic light below, of trivalent rare-earth ions in or near dielectric homogeneous spheres has been performed.
In the search for enhancement of fluorescence from magnetic light, one faces the difficult task of identifying the regions where the combined fluorescence due to multiple electric dipole (ED) transitions becomes negligible compared to the fluorescence of the MD transition.
We have succeeded in identifying a number of configurations with dielectric sphere parameters and a radial position of a trivalent rare-earth emitter wherein the branching ratio of the MD transition approaches its limit of one, implying that transitions from a given initial level (e.g., $^5$D$_0$-level of Eu$^{3+}$) are completely dominated by the MD transition.
The dimensionless directivity of the MD emission, the radiative decay rates, and the fluorescence of the magnetic light can be increased by a factor of more than $25$, $10^3$, and $10^4$, respectively.
\end{abstract}

\maketitle

\section{Introduction\label{sc:intr}}
Recent years have witnessed an increasing interest in generating light produced in magnetic dipole (MD) transitions~\cite{Li2017e,Baranov2017a,Sanz-Paz2018,Ernandes2018,Calandrini2018,Feng2018,Bidault2019,Darvishzadeh-Varcheie2019,Vaskin2019b,Wiecha2019,Yang2019b,Yang2019c,Wu2020a,Aslan2021,Aslan2022}.
We refer to this light as \textit{magnetic} light in the following.
Had the magnetic light been readily available, it would be possible to significantly expand and complement the present electromagnetic toolbox, which is based essentially only on the (electric) light generated in electric dipole (ED) transitions.
The principal obstacle for generating magnetic light is that the MD interaction with light is typically by the factor of the fine-structure constant ($1/137$) weaker than the ED interaction, and is mostly negligible.
Nevertheless, for certain quantum emitters, such as rare-earth ions~\cite{Deutschbein1939} for which the ED transitions (e.g., intraconfiguration transitions) can be forbidden~\cite{Walsh1998}, and semiconductor quantum dots~\cite{Cotrufo2015,Zurita-Sanchez2002}, the strength of MD transitions may be comparable or even greater than the competing ED ones.
Our main motivation is to enhance the magnetic light generation by rare-earth ions further beyond their conventional limits.

Similarly to ED transitions, a MD transition can be manipulated by engineering local photonic environments~\cite{Bhaumik1964,Snoeks1995,Karaveli2010,Karaveli2011,Taminiau2012,Aigouy2014,Kasperczyk2015,Feng2016a,Choi2016,Rabouw2016,Li2017e,Baranov2017a,Sanz-Paz2018,Ernandes2018,Wiecha2018,Calandrini2018,Feng2018,Bidault2019,Darvishzadeh-Varcheie2019,Mashhadi2019,Vaskin2019b,Wiecha2019,Yang2019b,Yang2019c,Zhong2019,Kalinic2020,Alaee2020a,Wu2020a,Sun2022}. 
A well-established tool for tailoring local photonic environments is provided by optical nanoantennas, which play central role for enhanced light-matter interactions in modern nanophotonics~\cite{Bharadwaj2009,Krasnok2013a}.
Integration of a quantum emitter into a nanoantenna that is capable of increasing the local density of optical states (LDOS) results in enhanced spontaneous emission rates. 
Given the relative weakness of magnetic transitions, the main focus in past decades has been placed on enhancement of ED transitions.
In the MD case, a mere optimization of decay rates has been considered so far~\cite{Klimov2005,Schmidt2012a,Karaveli2010,Karaveli2011,Rolly2012b,Taminiau2012,Kasperczyk2015,Zambrana-Puyalto2015,Feng2016a,Choi2016,Chigrin2016,Li2017e,Baranov2017a,Wiecha2018,Calandrini2018,Feng2018,Darvishzadeh-Varcheie2019,Mashhadi2019,Vaskin2019b,Wiecha2019,Zhao2019,Zhong2019,Kalinic2020,Alaee2020a,Habil2021,Hu2022MD}.
Several experimental works with ZrO$_2$~\cite{Cheng2021} and Si~\cite{Sugimoto2021} spheres, SiO$_2$ microcavities~\cite{DeDood2001,DeDood2001a}, and a number of numerical studies~\cite{Rolly2012b,Schmidt2012a,Zambrana-Puyalto2015,Chigrin2016,Zhao2019,Yao2021,Habil2021} (including ferromagnetic nanospheres~\cite{Neuman2020}) considered a number of different specific scenarios and concluded that either plasmonic or all-dielectric spheres are favourable for enhancing one of the ED and MD emission.
Other geometries have been considered as well: Si cuboids~\cite{Deng2022}, hollow Ge disk~\cite{Aslan2021}, Au nanocups~\cite{Mi2019}, and more complex structures~\cite{Pan2022,Aslan2022,Brule2022,Reynier2023,Puente2023}.

\begin{figure}
 \centering
 \includegraphics[width=3.33in]{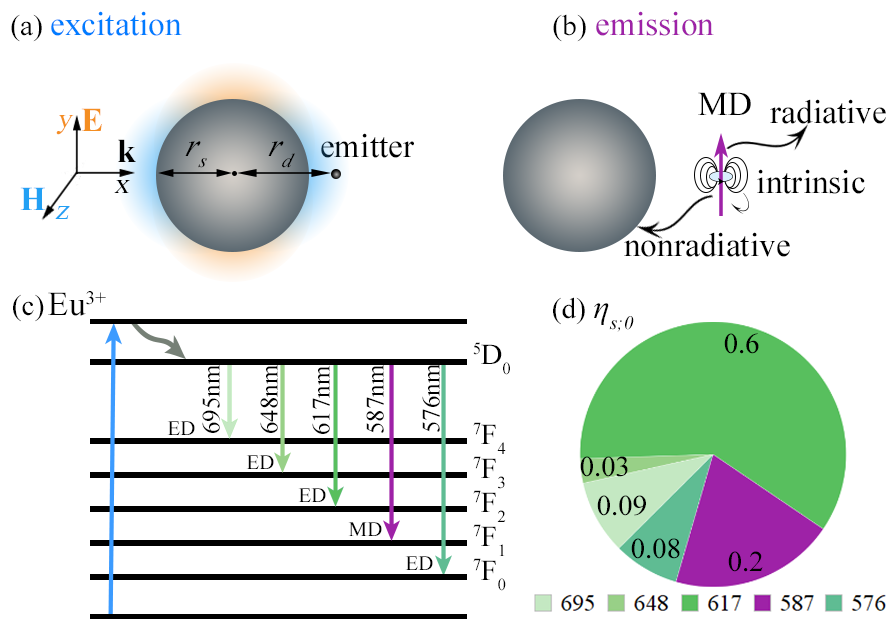}
 \caption{
 (a)-(b) Conventional two-step model for a particle-enhanced electric light fluorescence: 
 (a) excitation process under a plane wave illumination and 
 (b) emission process with intrinsic, radiative and non-radiative decay rates.
 (c) Simplified Jablonski diagram for Eu$^{3+}$~\cite{Yi2016,Luewarasirikul2021}.
 ${}^5$D${}_0 \to {}^7$F${}_5$ and ${}^5$D${}_0 \to {}^7$F${}_6$ transitions are typically broad and weak~\cite{Yi2016} and are not considered in this work.
 (d) Intrinsic branching ratios $\eta_{s;0}$ (in the absence of any particle) of the respective ${}^5$D${}_0 \to {}^7$F${}_s$ transitions of Eu$^{3+}$ in a dielectric matrix~\cite{Kolesnikov2016,Iwanaga2019,Shanmukha2022}.}
 \label{fig:scheme}
\end{figure}

Experimentally, amplifying MD emission is more difficult because most materials are not magnetic at optical frequency. There is no easy way to change the magnetic field and its associated properties. It is much easier to enhance ED because many materials have high permittivity enabling to manipulate the electric field.
In addition to the lack of suitable materials with appreciable magnetic permeability at optical frequencies, magnetic dipole transitions at optical frequencies are extremely rare and few elementary sources are available.
These sources exhibit typically a complex Jablonski diagram in which a magnetic dipole transition from a particular excited level does not occur in isolation, but is always accompanied by a series of different electric dipole transitions from the excited level (cf. Fig.~\ref{fig:scheme}(c)).
The latter makes magnetic light fluorescence fundamentally different from electric light fluorescence.
First, it cannot be described by a two-level approximation involving only magnetic dipole transitions.
Second, because magnetic dipole transition from a particular excited level does not occur in isolation, one faces a conundrum of resolving the competition of a magnetic transition from a set of different electric dipole transitions, all from a particular excited level (see Fig.~\ref{fig:scheme}(c)-(d)).
The latter requires to calculate for each individual transition its own radiative decay enhancement factor at the corresponding wavelength (see the Jablonski diagram in Fig.~\ref{fig:scheme}(c) and Table~\ref{tab:PL}).
Not surprisingly, complex aspects of magnetic light generation beyond rather straightforward decay engineering, such as affecting the MD/ED branching ratios, enhancing the fluorescence of magnetic light, and affecting its directionality, have been sparsely studied, if at all, even though the general theory for MD decay rates in the presence of a homogeneous sphere has been known for nearly twenty years~\cite{Klimov2005}.

In this work, we comprehensively deal with all the issues of magnetic light generation in rare-earth ions and present a detailed analysis of optimization of the MD emission, the MD/ED branching ratios, magnetic light fluorescence, and its directivity in the presence of an all-dielectric sphere. 
Some highlights to be discussed below involve: 
\begin{itemize}

\item[(i)] finding configurations at which the branching ratio of MD transition, $\eta_1$, can be substantially increased beyond its intrinsic value, and even approaching its limit value of unity;

\item[(ii)] observing that decay rates of magnetic light of trivalent rare-earth ions located inside or near dielectric homogeneous spheres can be enhanced by three orders of magnitude;

\item[(iii)] predicting directivity (the ratio of the radiation intensity in a given direction to the average radiation intensity in all directions) and fluorescence enhancement of MD emission as large as $\approx 26$ and $10^4$, respectively.
\end{itemize}

\section{Theory\label{sc:th}}

\subsection{Enhancement of magnetic light fluorescence}

The light-matter interaction between a charge-neutral quantum system and an electromagnetic field can be represented by a multipole expansion of the interaction Hamiltonian

\bg
H_{\rm int} = -\vp\cdot\vE(t) -\vm\cdot\vB(t)- [\vQ\vnab]\cdot\vE(t) -\ldots,
\label{Hint}
\eg
with $\vp$ being the electric dipole moment, $\vm$ the magnetic dipole moment, and $\vQ$ the electric quadrupole moment (a tensor).
It is obvious from $H_{\rm int}$ that the ED transitions are driven by electric field, $\vE$, whereas the MD transitions are driven by magnetic field, $\vB$.

The excitation and emission processes in MD fluorescence are in general separated by the Stokes shift. 
Each of the respective processes is described by its corresponding part of $H_{\rm int}$.
In principle, both ED and MD transitions can be used to promote an occupation level of MD transitions~\cite{Kasperczyk2015}.
Rare earth ions emission have been known as early as since 1939 to comprise magnetic light~\cite{Deutschbein1939,Judd1962,Ofelt1962,Linares1966}.
In a trivalent rare-earth the emission level (e.g., $^5$D$_0$ for Eu$^{3+}$) can be populated by an initial excitation to a higher $d$-level (e.g., $^5$D$_1$). 
The initial excitation may involve either ED transitions, MD transitions, or both.
For example, the $^5$D$_0$ level of Eu$^{3+}$ decays radiatively through either ED transitions [${}^5$D${}_0\to {}^7$F${}_j$ ($j=0,2,3,\ldots, 6$)] or the MD transition (${}^5$D${}_0\to {}^7$F${}_1$)~(see Fig.~(1) in Ref.~\cite{Rikken1995}).
The ED (MD) transitions can be promoted with \textit{electric} (\textit{magnetic}) Mie resonances~\cite{Rolly2012b}, which is formally described by $H_{\rm int}$ of Eq.~(\ref{Hint}). 

As the above example of Eu$^{3+}$ shows, any MD fluorescence of trivalent rare-earth will be accompanied by ED fluorescence.
To characterize the relation between MD and ED fluorescence, we introduce the branching ratio of the MD transition as~\cite{Ofelt1962,Walsh1998,Nagaishi2003,Luewarasirikul2021}
\bg
\eta_s=\dfrac{\Gm_{{\rm rad};s}}{\sum_j \Gm_{{\rm rad};j}},
\label{brdf}
\eg
where $\Gm_{{\rm rad};j}$ is the radiative decay rate of $j$-th transition and $j=s$ corresponds to a MD transition (e.g., $s=1$ in the case of $^5$D$_0$ level of Eu$^{3+}$ shown in Fig.~\ref{fig:scheme}).
The emission strength of MD transition and the multitude of ED transitions is determined by the selection rules.
Improving magnetic light fluorescence therefore presents a challenge in improving the branching ratio of MD compared to the numerous ED transitions as presented in our Eq.~(\ref{brdf}).
Therefore any search for enhancement of fluorescence from magnetic light presents difficult task of identifying the regions where the combined fluorescence due to multiple electric dipole (ED) transitions becomes negligible compared to the fluorescence of the MD transition.

In what follows we shall make use of \textit{orientation averaged} radiative decay rates in the presence of a sphere that are entering the branching ratio in Eq.~(\ref{brdf})~\cite{Moroz2005,Bharadwaj2007}: $\Gm_{{\rm rad};j} = (\Gm^\perp_{{\rm rad};j} + 2\Gm^\parallel_{{\rm rad};j})/3$, where superscripts ``$\perp$'' and ``$\parallel$'' denote the radial and tangential orientation of the dipole emitter with respect to a sphere surface.

There are several critical features distinguishing typical MD fluorescence from a conventional two-step ED fluorescence within a two-level approximation:
\begin{itemize}

    \item The excitation paths, and thus the respective wavelengths, dramatically vary with the experimental setup, as shown in Table~\ref{tab:MD_ions} for Eu$^{3+}$ ions. 
    At the same time, either magnetic~\cite{Rikken1995,Sugimoto2021a} or electric field enhancement can be used to excite triplet state.
   
    \item One parent level (e.g., $^5$D$_0$ level of Eu$^{3+}$ shown in Fig.~\ref{fig:scheme}) may radiatively decay into a set of $N$ daughter levels (e.g., $^7$F$_j$ levels, $j=0,1,\ldots 6$, of Eu$^{3+}$ shown in Fig.~\ref{fig:scheme}). 
    Unless the branching ratio $\eta_s$ defined by Eq.~(\ref{brdf}) is changed, the excitation enhancement will equally enhance all the transitions.
    Therefore the main goal of the sphere is to exert control over emission processes.

\end{itemize}
\begin{table}
    \centering
    \caption{Summary of the available data for experimentally measured emission wavelength $\ld_{\rm ems}$ of the MD transition ${}^5$D${}_0\to {}^7$F${}_1$ of Eu$^{3+}$ ions in different solid hosts. $\ld_{\rm exc}$ denotes corresponding excitation wavelength.}
    \vst{0.3cm}
    \begin{tabular}{c|c|c|c}
         configuration & $\ld_{\rm exc}$~(nm) & $\ld_{\rm ems}$~(nm) & Reference \\
          &&& \\
         \hline 
          &&& \\
         Eu$^{3+}$:in solution & 527.5 & 590 & \cite{Deutschbein1939,Freed1941,Kasperczyk2015}  \\
         Eu$^{3+}$:TiO$_2$ on Si & 325/442 & 593.4 &  \cite{Conde-Gallardo2001} \\
         Eu$^{3+}$:TiO$_2$ on glass & 325/442 & 593.1 & \cite{Conde-Gallardo2001} \\
         Eu$^{3+}$:TiO$_2$ on Si & 325 & 593.9 & \cite{Peng2005} \\
         Eu$^{3+}$:TiO$_2$ & 468.3 & 595 & \cite{Antic2012} \\
         Eu$^{3+}$:TiO$_2$ & 250 & 588 & \cite{Reszczynska2016} \\
         Eu$^{3+}$:SiO$_2$(nanowires) & 393 & 592 & \cite{Gao2022} \\
         Eu$^{3+}$:SnO$_2$ & 310 & 593 & \cite{Liu2011b} \\ 
         Eu$^{3+}$:Bi$_2$SiO$_5$ & 394 & 595 & \cite{Chen2021b} \\ 
         Eu$^{3+}$:Ag:G & 464 & 592 & \cite{Wei2012} \\
         Eu$^{3+}$:Si & 405 & 590 & \cite{Sugimoto2021} \\
         Eu$^{3+}$:ZrO$_2$ & 325 & 587 & \cite{Cheng2021} \\
         &&& \\
         \hline 
    \end{tabular}
    \label{tab:MD_ions}
\end{table}
Due to the above reasons, earlier formulas used within the two-level approximation~\cite{Bharadwaj2007,Sun20JPCC,Rasskazov21JPCL} cannot be employed here and their adaptation to the present multilevel case is required.
Our key assumption is that the single-parent level undergoes a single-exponential decay into a number of daughter levels. 
Consequently, the measured fluorescence signal follows an exponential law,
\bg
I_{f}(t) = I_{f}(0)\,  e^{-t \Gm_{\rm tot}},
\label{sel}
\eg
where $I_{f}(0)$ is the initial fluorescence signal at time $t=0$.
$\Gamma_{\rm tot}$ is the sum over all possible decay rates from the parent level, $\Gamma_{\rm tot}=\sum_j (\Gm_{{\rm rad};j}+\Gm_{{\rm nrad};j}+\Gm_{{\rm int};j})$, where $\Gm_{{\rm \cdot};j}$ are the respective (radiative, nonradiative and ``intrinsic'' nonradiative in the absence of any particle) rates for the decay from the single parent level (e.g., in the case of $^5$D$_0$ level of Eu$^{3+}$ shown in Fig.~\ref{fig:scheme}) into $j$-th daughter level (e.g., the case of $^7$F$_j$ levels of Eu$^{3+}$ shown in Fig.~\ref{fig:scheme}).

The population of the parent level, $N_r$, i.e., the number of excited sources, decreases during an infinitesimal time interval ${\rm d}t$ as 
\bg
{\rm d}N_r = -\Gm_{\rm tot} N_r(t)\, {\rm d}t.
\eg
After switching off the excitation at $t=0$, the total detected fluorescent intensity, $I_{\rm tot}$, shall be proportional to  the number of excited sources $N_r$,
\bg
I_{\rm tot}=\int_0^\infty I_{f}(t)\,   {\rm d\textit t} = g N_r(0),
\label{itot}
\eg 
where $g$ is a proportionality factor to be determined below. 
Depending on the kind of experiment, $I_{f}$ can have the meaning of fluorescence count in Hz. 
Within the single exponential decay of Eq.~(\ref{sel}), $I_{\rm tot}$ is, by performing the integral in Eq.~(\ref{itot}), inversely proportional to the total decay rate $\Gm_{\rm tot}$,
\bg
I_{\rm tot}=\int_0^\infty  I_{f}(0) e^{-t \Gm_{\rm tot}}\, {\rm d}t= \fr{I_{f}(0)}{\Gm_{\rm tot}}\cdot
\label{flt}
\eg
It is expedient to define the corresponding fluorescence quantum efficiencies,
\bg
q_j = \fr{\Gm_{{\rm rad};j}}{\Gm_{\rm tot}}\cdot
\label{etad}
\eg
Each fluorescence quantum efficiency $q_j$ is essentially the probability of a radiative transition to the $j$-th daughter level.
Thus, not surprisingly, after switching off the excitation at $t=0$, the total detected fluorescent intensity is
\bg
I_{\rm tot}  = \left(\sum_j q_j\right) N_r(0),
\label{fl2}
\eg
which determines the proportionality constant $g$ in Eq.~(\ref{itot}).
Given that $N_r(0)$ is proportional to the excitation rate, Eqs.~(\ref{etad}) and~(\ref{fl2}) constitute the required generalization of a two-level fluorescence to the case of a single parent level decaying into a plurality of daughter levels.
For $j=1$, one recovers the usual formulas o,f e.g., Refs.~\cite{Bharadwaj2007,Sun20JPCC,Rasskazov21JPCL}.
Obviously, one has to have
\bg
I_{f}(0) =\left(\sum_j \Gm_{{\rm rad};j} \right) N_r(0)
\label{fl1}
\eg
in order for Eq.~(\ref{flt}) to be compatible with Eq.~(\ref{fl2}).
Equation~(\ref{fl1}) could in turn be used to determine the total radiative rate ($\sum_j  \Gm_{{\rm rad};j}$) experimentally. 
  
The presence of a particle modifies the LDOS (cf. Fig.~\ref{fig:scheme}), in which case each $q_j$ in Eq.~(\ref{fl2}) becomes a nontrivial function of emitter position.
Obviously, at the spatial infinity is each $q_j$ approaching its value in the absence of a particle.
Enhancement of $\eta_1$ by a sphere in the case of the MD transition into $^7$F$_1$ level of Eu$^{3+}$, shown in Fig.~\ref{fig:scheme}, will lead to a higher proportion of the MD fluorescence.
To tilt the natural branching ratios in favor of MD emission, it is expedient to tune the sphere resonance to the MD emission path.
In the following, we investigate to what extent it is possible to promote MD fluorescence compared to ED fluorescence by the presence of a dielectric homogeneous spherical particle (see Fig.~\ref{fig:scheme}), assuming excitation by plane waves.

According to Fig.~\ref{fig:scheme}(d), the ``intrinsic'' (i.e., in the absence of any particle) MD branching ratio $\eta_{1;0}\approx 0.2$.
Experimental determination of radiative rates $\Gm_{{\rm rad};j}$ is known to be tricky.
Semi-empirical Judd-Ofelt theory~\cite{Judd1962,Ofelt1962,Walsh1998} enables one to approximate radiative lifetimes and determine the branching ratio of rare-earth in different solid crystal matrices.
Nevertheless, experimentally verification of the Judd-Ofelt parametrization is typically not better than within  $15-30\%$ of the theoretical values~\cite{Walsh1998}. 
For the sake of illustration, we take the initial free space values of the respective radiative rates $\Gm_{{\rm rad};j}$ as those proportional to corresponding photoluminescence (PL) intensities (defined in the arbitrary units) summarized in Table~\ref{tab:PL}. 
\begin{table*}[hbt!]
\caption{Photoluminescence intensities of refs \cite{Kolesnikov2016,Iwanaga2019,Shanmukha2022} used to determine the branching ratios of Eu$^{3+}$ emission in Fig.~\ref{fig:scheme}(d) and respective absolute values of free space radiative rates.}
    \centering
    \vst{0.4cm}
    \begin{tabular}{c|c|c|c|c|c }
        transition & ${}^5$D${}_0 \to {}^7$F${}_4$ & ${}^5$D${}_0 \to {}^7$F${}_3$ & ${}^5$D${}_0 \to {}^7$F${}_2$ & ${}^5$D${}_0 \to {}^7$F${}_1$ & ${}^5$D${}_0 \to {}^7$F${}_0$ \\ 
        &&&&& \\
         \hline 
          &&&&& \\
        nature & ED & ED & ED & MD & ED \\ 
        $\ld_{\rm ems}$~(nm) & 695 & 648 & 617 & 587 & 576 \\
        PL~(arb. un.) & 37.1 & 13.6 & 241.3 & 81.4 & 31.1 \\
        $\Gm^0_{\rm rad;j}$~(s$^{-1}$) & 81.4 & 29.9 & 530.9 & 179.1 & 68.4 \\
        &&&&& \\
        \hline 
    \end{tabular}
    \label{tab:PL}
\end{table*}

In order to turn the values into sensible absolute radiative rates, one has to sum up all the PL intensities and compare the result with the known total decay rate of the excited state of trivalent rare earth, $\Gm_R=891$ s$^{-1}$ for Eu$^{3+}$~\cite{Lima2016}. 
This way, a normalization factor can be established to translate the PL intensities into the absolute radiative rates.
In order to get position-dependent absolute values of $\Gm_{{\rm rad};j}$ in the presence of a spherical particle, each initial free space value of $\Gm_{{\rm rad};j}$ in Table~\ref{tab:PL} is multiplied with its corresponding enhancement factor at the corresponding transition wavelength. 
Those values are then substituted into Eq.~(\ref{brdf}), thereby obtaining position-dependent branching ratios.

Finally, the ultimate fluorescence enhancement factor for $j$-th transition can be recast as: 
\bg
F_j = \dfrac{\gm_{\rm exc}}{\gm_{\rm exc;0}}\dfrac{q_j}{q_{j;0}} \ ,
\label{f}
\eg
where $\gm_{\rm exc}$ is the excitation rate and subscript ``0'' denotes the respective quantity in the absence of particle.

\subsection{Duality symmetry of the Maxwell's equations in the dipole case}
According to Ref.~\cite{Jackson1999}~(see Chapter 9, Page 414), the behavior of fields $\vE_{\rm md}$ and $\vH_{\rm md}$ for a magnetic dipole source in the near and far zones are the same as for the fields $\vE_{\rm ed}$ and $\vH_{\rm ed}$ of the electric dipole source, with the interchange ($\vE_{\rm md}$ and $\vH_{\rm md}$ in Eq.~(\ref{emdds}), see Appendix~\ref{sc:dpemsym} for the derivation)
\bg 
\vE_{\rm ed} \to Z \vH_{\rm md},\quad Z\vH_{\rm ed} \to - \vE_{\rm md},\quad \vp\to \vm/c,
\label{J414}
\eg 
where $Z=\sqrt{\mu/\veps}$ is the impedance of the medium hosting the dipole, $\vp$ and $\vm$ are electric and magnetic dipole moments, respectively.
In the Gauss units, the duality symmetry implies $\vE_{\rm ed} \to \vB_{\rm md}$, $\vB_{\rm ed} \to -\vE_{\rm md}$, $\vp\to \vm$.
Similarly, the radiation pattern and total power radiated are the same for the two kinds of dipoles.
The only difference in the radiation fields is in the \textit{polarization}.
For an electric dipole, the electric field [$\sim (\vn\times\vp)\times \vn=\vp-(\vp\cdot\vn)\vn$] lies in the plane defined by $\vn$ and $\vp$, while for a magnetic dipole the electric field [$\sim (\vn\times\vm)$] is perpendicular to the plane defined by $\vn$ and $\vm$ (see page 414 in Ref.~\cite{Jackson1999}).

The symmetry of the problem described by Eq.~(\ref{J414}) enables one to translate the known results for the relative radiative and nonradiative decay rates of an ED emitter in the presence of a sphere to those of an MD emitter by simply swapping transverse electric (TE) and transverse magnetic (TM) mode labels in respective equations for dipole emitter (see Eq.~(124-134) in Ref.~\cite{Moroz2005} and Eq.~(28) in Ref.~\cite{Rasskazov20OSAC}).
Note in passing that the TM (TE) modes are also known as \textit{electric} (\textit{magnetic}) modes.
Respective equations for both ED and MD emitters are presented in Appendix~\ref{sc:dpemsym}.

The required task of determining relevant fluorescence parameters is much more involved than in the previous studies involving electric dipole fluorescence~\cite{Bharadwaj2007,Sun20JPCC,Rasskazov21JPCL}, because it necessitates not only calculation of enhancement factors for both magnetic dipole and electric dipole transitions, but also determining for each $\Gm_{{\rm rad};j}$ its own enhancement factor at its corresponding wavelength according to the Jablonski diagram in Fig.~\ref{fig:scheme} (see Table~\ref{tab:PL}). 

Simulations presented below are performed by freely available MATLAB code Stratify~\cite{Rasskazov20OSAC} after implementing the above symmetry properties therein.
For the rest of the paper, the spherical particle is embedded in a homogeneous medium with a refractive index $n_h$, which is set to be air ($n_h=1$).

\section{Results and discussion}

\subsection{Single Emitter}
As an experimentally accessible example, we consider TiO$_2$ spheres with the real refractive index $n_s=2.7$ and zero imaginary part, which is a valid approximation~\cite{Baranov2017b}.
An emitter will be considered at a variable radial position inside and outside a sphere. 
From an experimental point of view, rare earth can be embedded into spheres with nanometer precision in radial direction for a number of different materials~\cite{VanBlaaderen1992,Gritsch2022}.
Controlled positioning of emitters outside the sphere is possible by attaching emitters to (nano)particles via single-stranded DNA (ssDNA) spacers~\cite{Dulkeith2005} and DNA origami~\cite{Acuna2012}.

\begin{figure*}
 \centering
  \includegraphics[width=6in]{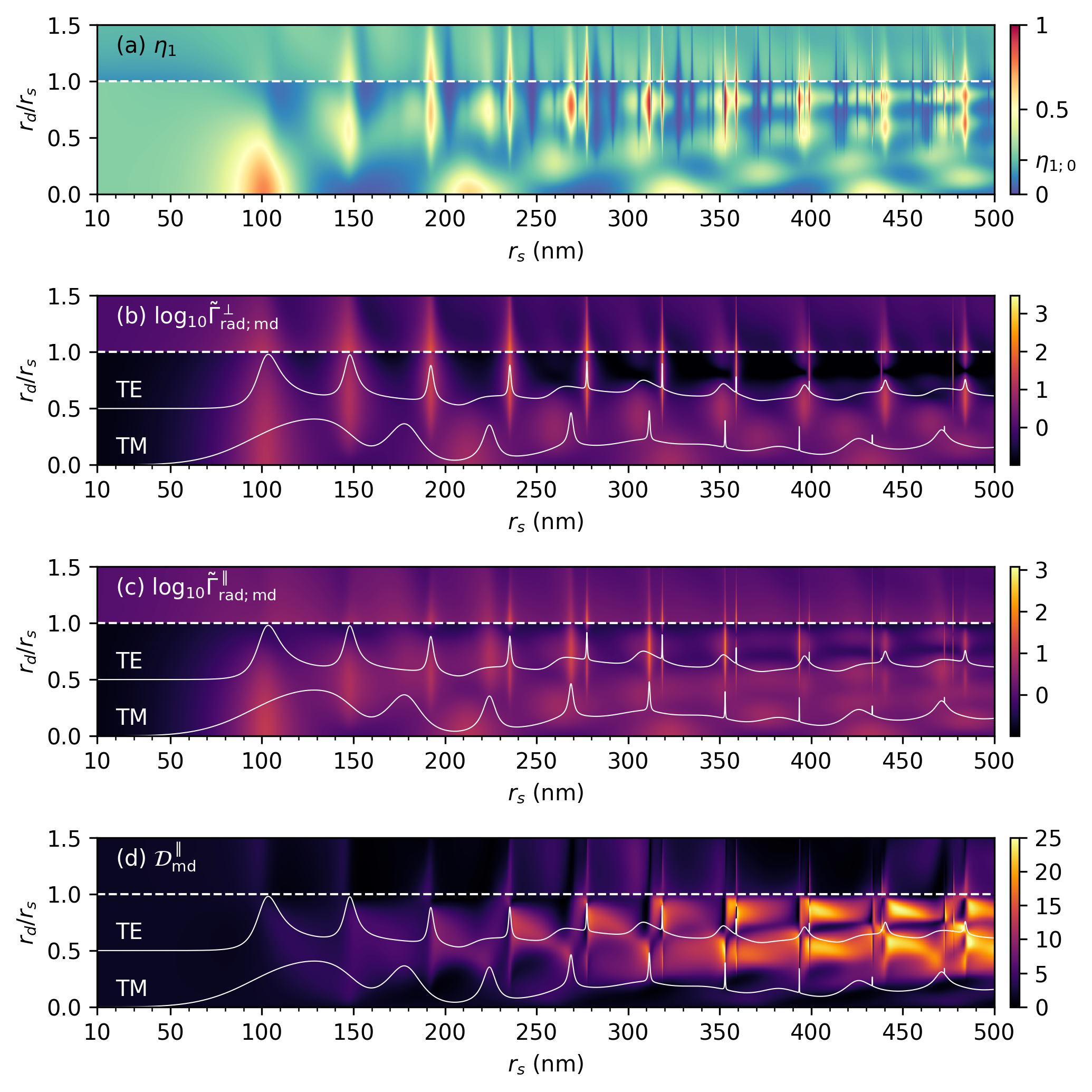}
  \caption{Color maps of the
  (a) branching ratio, 
  (b) the radial ($\perp$) and 
  (c) tangential ($\parallel$) components of MD radiative decay rates, and 
  (d) the directivity $\mathcal{D}^{\parallel}_{\rm md}$ of emission of the tangentially oriented MD emitter with $\ld_{\rm ems;md}=587$~nm (see Fig.~\ref{fig:scheme}) as a function of normalized dipole position, $r_d/r_s$, and TiO$_2$ sphere ($n_s=2.7$) radius $r_s$. 
  The horizontal dashed lines at $r_d/r_s = 1$ denote the surface of a sphere and distinguish the location of emitters inside ($r_d < r_s$) and outside ($r_d > r_s$) sphere. 
  The directivity of the emission of the tangentially oriented MD (see Fig.~\ref{fig:scheme}(a)) is shown in (d).
  The directivity of radially oriented MD (not shown) is close to zero for any wavelength and $r_s$.
  Solid white lines in (b)-(d) show the extinction of corresponding TiO$_2$ spheres for the TM and TE polarizations (with an offset of 0.5 for clarity) at $\ld_{\rm ems;md}=587$~nm. 
  Note pronounced correlations of the radially (b) oriented MD emission properties with the peaks of the \textit{magnetic} (TE) resonances and of the tangentially (c) oriented MD emission properties with the peaks of both \textit{magnetic} (TE) and \textit{electric} (TM) resonances.}
 \label{fig:MD_TiO2}
\end{figure*}
In what follows, we focus first on finding regimes with simultaneously large $\eta_1$ and $\tilde\Gm_{\rm rad;md} = \Gm_{\rm rad;md} / \Gm_{\rm rad;md;0}$, where subscript ``0'' denotes the respective quantity in the absence of particle.
Figure~\ref{fig:MD_TiO2}(a) shows that the MD branching ratio approaches the limit value $\approx 1$ for a variety of sphere sizes and emitter positions.
As obvious from Eq.~(\ref{brdf}), $\eta_1= 1$ is the largest value which the branching ratio for MD transition can ever attain.
Therefore, for the configurations at which $\eta_1$ approaches unity, the transitions from the parent level $^5$D$_0$ level of Eu$^{3+}$ are entirely dominated by the MD transition.
Figure~\ref{fig:MD_TiO2} shows close correlations between peaks of the branching ratio, MD radiative decay rates, and the directivity $\mathcal{D}_{\rm md}$ of emission of (tangentially oriented) MD with the \textit{magnetic} (TE) resonances.
Directivity $\mathcal{D}_{\rm md}$ is a relation of the power emitted into a certain direction (in our case, into the $-x$ direction) to the solid angle averaged emitted power see Eq.~(\ref{Ddf}) in Appendix~\ref{sc:directivity} for details.
For the parallel dipole orientation, magnetic and electric transitions have decay rate enhancement factors that are spectrally well separated.
For the perpendicular dipole orientation, both electric and magnetic dipole emitters exhibit common maxima, consistent with Ref.~\cite{Rolly2012b}.
Indeed, according to Eqs.~(\ref{eq:EDdecay}) and~(\ref{eq:MDdecay}) (see Appendix~\ref{sc:mdrates}), one may expect radiative decay rates for the radially oriented ED and MD emitters to follow the TM and TE resonances of spheres, respectively.
At the same time, the tangentially oriented ED and MD emitters exhibit enhanced radiative decay rates at both TE and TM resonances.

\begin{figure}[t!]
 \centering
  \includegraphics[width=3.33in]{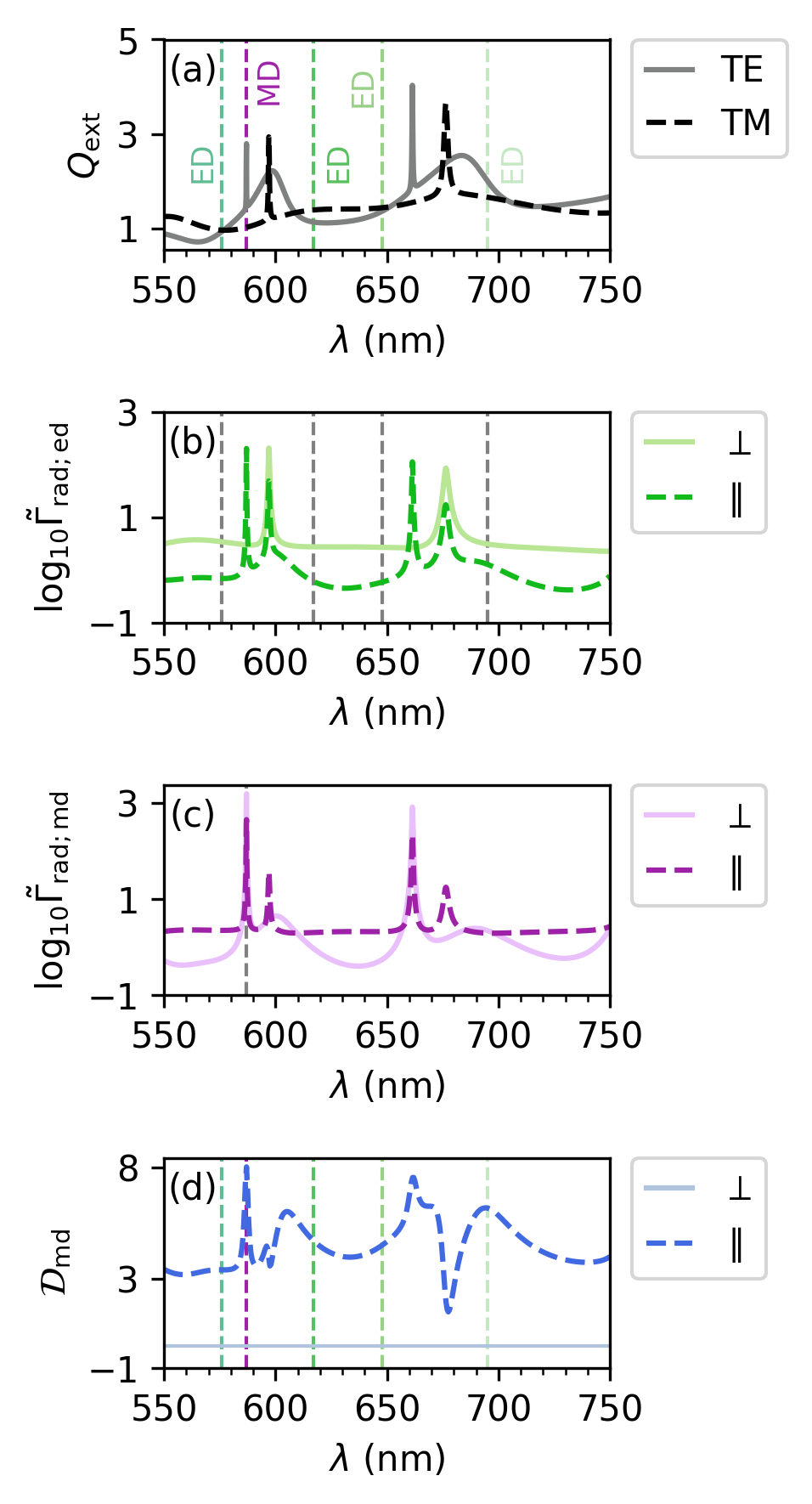}
  \caption{
  (a) Extinction efficiency,
  (b) ED radiative decay rates,
  (c) MD radiative decay rates for the respective radial ($\perp$) and tangential ($\parallel$) Eu$^{3+}$ emitter orientations, and
  (d) directivity $\mathcal{D}_{\rm md}$ of the MD emission, all as a function of wavelength in the presence of TiO$_2$ sphere in air ($n_h=1$) of fixed radius $r_s=359$~nm. The sphere size provided for the maximum branching ratio of the MD channel in Fig.~\ref{fig:MD_TiO2}(a). The
MD emitter position is fixed at $r_d=r_s+1$~nm.
  Dashed vertical lines indicate the respective ED (MD) ${}^5$D${}_0 \to {}^7$F${}_j$ transitions of Eu$^{3+}$ emitter (see Fig.~\ref{fig:scheme}(c)).}
 \label{fig:TiO2ext}
\end{figure}
\begin{figure}[hbt!]
 \centering
 \includegraphics[width=3.33in]{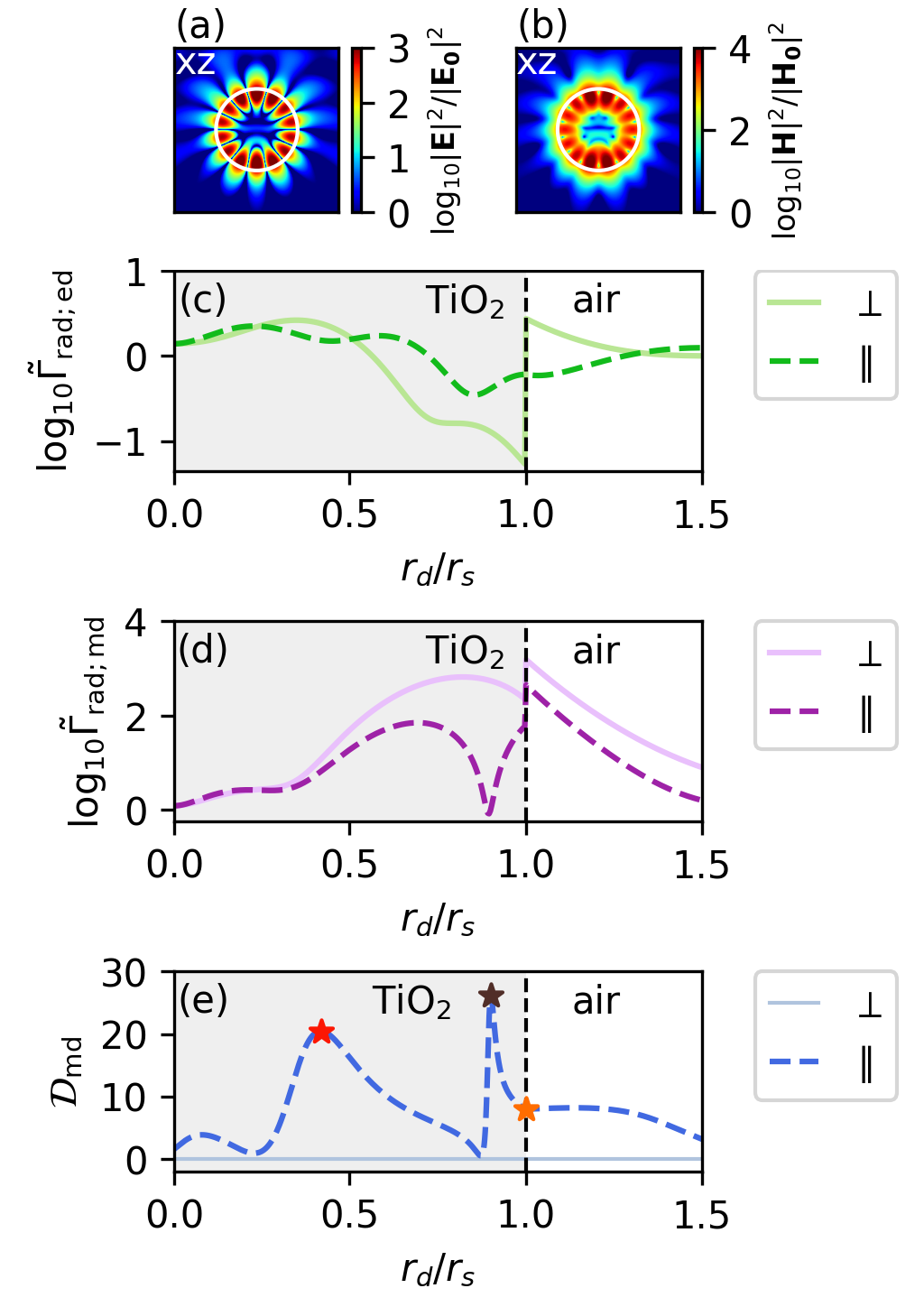}
 \caption{
 (a) Electric, $\bf E$, and 
 (b) magnetic, $\bf H$, local field distribution near TiO$_2$ sphere of fixed radius $r_s=359$~nm at $\ld_{\rm ems;md}=587$~nm in air ($n_h=1$).
 Incident plane-wave amplitude $\vE_0$ oscillates along the $y$ axis and propagation is along the $x$ axis as indicated in Fig.~\ref{fig:scheme}(a).
 (c)-(d) Radiative decay rates as a function of emitter radial position $r_d/r_s$ for the radially ($\perp$) and tangentially ($\parallel$) oriented 
 (c) ED and (d) MD.
 (e) Respective directivities $\mathcal{D}_{\rm md}$ for the two orthogonal orientations of the MD emitter as a function of emitter radial position $r_d/r_s$ for the same TiO$_2$ sphere. 
 Vertical dashed lines in (c)-(e) at $r_d/r_s = 1$ denote the surface of a sphere and distinguish the location of emitters inside ($r_d < r_s$) and outside ($r_d > r_s$) sphere.
 Emission wavelengths of the highlighted ED and MD transitions are different and correspond to $\ld_{\rm ems;ed}=617$~nm and $\ld_{\rm ems;md}=587$~nm, respectively, as exhibited in Fig.~\ref{fig:scheme}(c).
 The ED transition at $\ld_{\rm ems;ed}=617$~nm has been chosen due to its dominant intrinsic branching ratio.
 ED transitions at other wavelengths of Fig.~\ref{fig:scheme}(c) demonstrate similar radiative decay rates (not shown).
 Stars in (e) correspond to the cases considered in detail in subsequent Fig.~\ref{fig:far}.
 }
 \label{fig:TiO2rad}
\end{figure}

To get more insight, Fig.~\ref{fig:TiO2ext} shows wavelength-dependent extinction of TiO$_2$ sphere with $r_s=359$~nm radius, decay rates, and directivity of Eu$^{3+}$ emitter located on the surface of such a sphere.
According to Fig.~\ref{fig:MD_TiO2}, this corresponds to $\eta_1\approx 1$ and fairly large $\tilde\Gm^{\perp,\parallel}_{\rm rad;md}\approx 10^3$.
It can be seen from Figs.~\ref{fig:TiO2ext}(a-c) that the possible ED transitions of Eu$^{3+}$ emitter do not correlate with any of resonances.
Contrary to that, the MD transition of Eu$^{3+}$ matches the magnetic (TE) resonance of a sphere.
This is the physical origin behind dominant MD emission in this case.
As a pleasant bonus, the MD emission for the tangentially oriented MD emitter has relatively high directivity $\cal{D}_{\rm md}$.
\begin{figure*}[hbt!]
 \centering
 \includegraphics[width=5in]{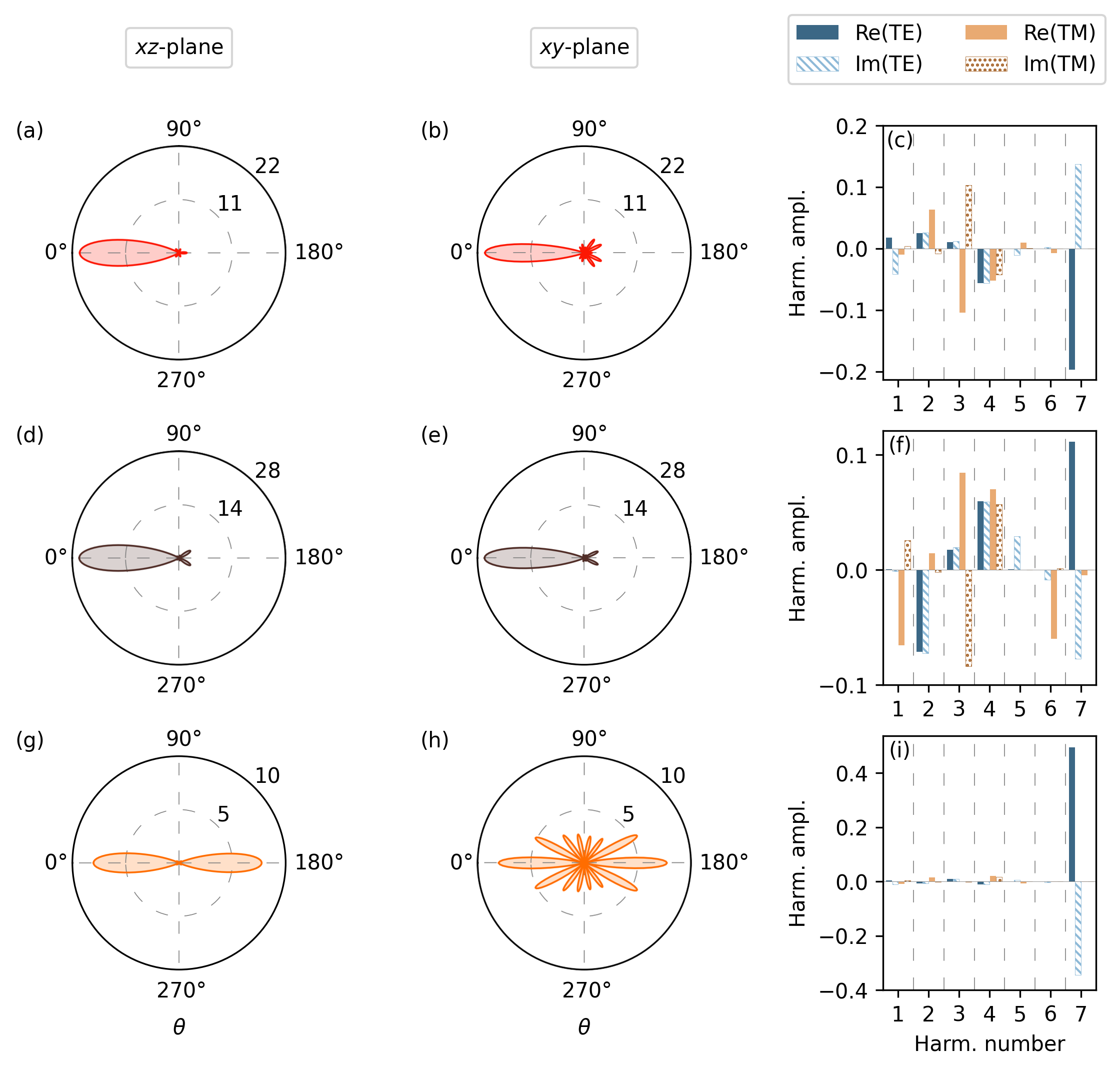}
 \caption{Polar plots of the angle-resolved directivity $\mathcal{D}_{\rm md}$ for the $xz$-plane (left column) and $xy$-plane (middle column) of the Eu$^{3+}$ MD emission and their respective multipolar decomposition (right column) for the tangentially oriented emitters at the radial locations highlighted in Fig.~\ref{fig:TiO2rad}(e) (from left to right):
 (a)-(c) $r_d/r_s=0.418$,
 (d)-(f) $0.901$,
 (g)-(i) $1$.  
 Polar angle $\theta=0^{\circ}$ indicated in the figures corresponds to the direction opposite to the $x$-axis. The emitter is assumed to be on the $x$-axis and oriented along the $y$-axis as shown in Fig.~\ref{fig:scheme}(a).
 The amplitudes of the harmonics are normalized to the sum of the absolute values of the amplitudes for all excited harmonics.
 Numbers on concentric circles in the polar plots reflect corresponding directivity values.}
 \label{fig:far}
\end{figure*}

Electromagnetic fields for the TE resonances at $\lambda=587$~nm corresponding to the MD transition are plotted in Figs.~\ref{fig:TiO2rad}(a,b). 
The electric and magnetic local field distribution near TiO$_2$ sphere enables one to identify this resonance as a multipolar $l=7$ resonance.
In  Figs.~\ref{fig:TiO2rad}(c,d) it is shown how the radiative decay rate of Eu$^{3+}$ depends on the position of the MD emitter inside or outside TiO$_2$ sphere.
It is worth noting that the maximum achievable radiative decay rates (for both the ED and MD transitions) occur when the emitter is located near the surface of the sphere. 
According to Figs.~\ref{fig:TiO2rad}(c,d), the MD radiation dominates the ED radiation by several orders of magnitude for any radial position of emitter.

Interestingly, the directivity $\mathcal{D}_{\rm md}$ of such a system attains several maxima when the MD emitter is inside the particle, and a relatively large value when the emitter is near its surface. 
For a complete understanding of the formed radiation pattern in the far zone, an expansion in terms of electric (TM) and magnetic (TE) spherical harmonics was performed, indicating their amplitudes and phases.
Figure~\ref{fig:far} shows the dependencies of the directivity $\mathcal{D}_{\rm md}$ of a tangentially oriented MD emitter inside a TiO$_2$ sphere in air ($n_h=1$) on the polar black angle $\theta$ for the cases marked in Fig.~\ref{fig:TiO2rad}(e) presented in the same order. 
Polar angle $\theta=0^{\circ}$ corresponds to the direction along the negative $x$ axis.
The data are presented as polar plots in the $xz$ plane (left column) and $xy$ plane (central column), and also the corresponding multipolar decomposition (right column).
It is well known that the stored energy of different modes of a spherical resonator is concentrated in certain regions~\cite{Bohren1986}.
A change in the position of the magnetic dipole of a spherical resonator has a different effect on the excitation of different modes of a given resonator. Nevertheless, the energy is usually concentrated inside the dielectric.
Therefore, when the dipole is placed inside, it becomes possible to adjust the amplitudes and phases of the different harmonics in such a way that all side lobes in the radiation pattern are suppressed or significantly reduced.
Notably, angle-resolved directivities inside and outside the sphere are remarkably different: inside the sphere the superposition of induced multipoles leads to highly directive emission in the forward direction (Figs.~\ref{fig:far}(a,b,d,e)), while for the emitter located outside sphere, the dominant $l=7$ mode gives rise to a symmetrical radiation in the forward and backward directions (Figs.~\ref{fig:far}(g,h)).
Note in passing that the directivity of the MD emission for the tangentially oriented emitters can achieve $D_{\rm md}\approx 26$ for the described configuration.
A Fano-type behavior of the directivity in Fig.~\ref{fig:TiO2rad}(e) can be explained by mixing of different multipolar modes with different amplitudes and phases induced in the sphere by the MD, as shown in Figs.~\ref{fig:far}(c,f,i).
These expansions take into account both the radiation from the excited modes of the dielectric resonator and from the magnetic dipole itself.

\begin{figure*}
 \centering
  \includegraphics[width=6in]{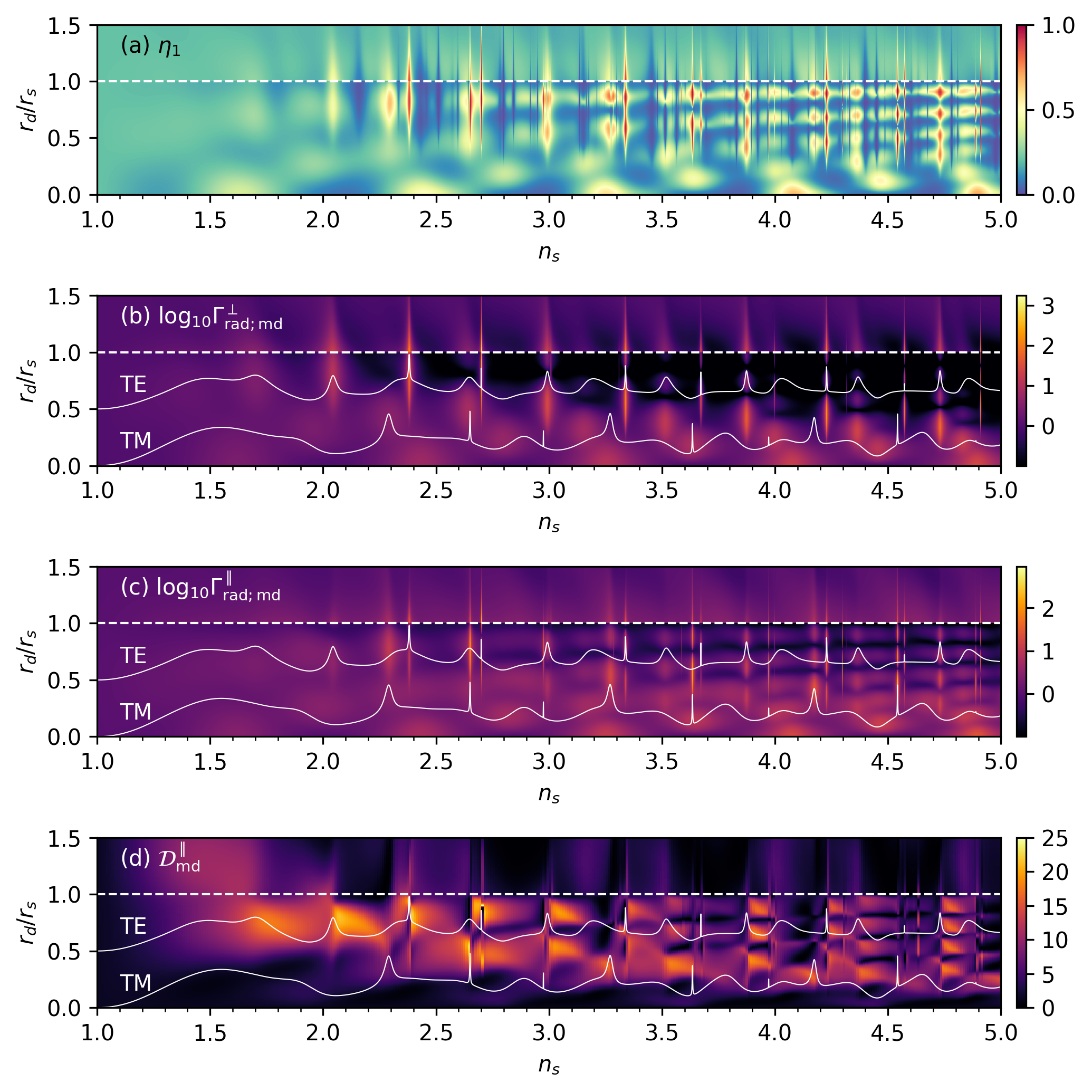}
  \caption{Same as in Fig.~\ref{fig:MD_TiO2}, but for a sphere with fixed radius $r_s=359$~nm in air ($n_h=1$) as a function of sphere refractive index $n_s$.}
 \label{fig:MD_ns}
\end{figure*}

Figure~\ref{fig:MD_ns} demonstrates that our results can be translated to spheres with other refractive indexes $n_s$ 
and to other wavelength ranges, taking into account the respective intrinsic branching ratios and the relevant variety of ED and MD transitions of emitters under consideration.
The above scalability ensures wide applicability of the presented results for magnetic light generation.
For instance, our work could serve as an important tool for vibrational circular dichroism (VCD)~\cite{Stephens1985,Nafie2004}, since the very chiral signatures are attributed to the MD transitions, which enhancement is critical for sensing applications~\cite{Ho2017}.
In this regard, recently emerged high-index iron pyrite~\cite{Doiron2022} with nearly zero losses and impressively large $n_s\leq4.47$ in mid-IR range could be a very promising sphere material.

\subsection{Multiple Emitters}
The effect of multiple rare earth emitters on the presented above results deserves a thorough discussion.
It has been known that a finite density of rare earth emitters results in the so-called \textit{concentration quenching}~\cite{Linares1966,DeDood2001,Dulkeith2005}.
Once an emitter is excited, it can transfer its excitation to a nearby emitter via an energy transfer mechanism.
With multiple emitters present, this transfer process may end up with an eventual decay of an emitter excitation.
However, if the excitation is transferred to an emitter in a proximity of an impurity (e.g., OH bonds in silica), the excitation can be swallowed up by the impurity, whereby the excitation disappears without ever contributing to either the ED or MD radiation.
This is the essence of the concentration quenching.
Obviously, the concentration quenching contributes to the nonradiative decay rate which origin is different from the Ohmic losses.
The \textit{non-radiative} decay rate is described by
\bg
{\Gm}_{\rm nrad} = 8\pi C_{\rm Eu-Eu}[{\rm Eu}^{3+}][Q] \ ,
\eg
where [Eu$^{3+}$] ([$Q$]) is the emitter (quencher) concentration in at. \%, and $C_{\text{Eu-Eu}}$ is a coupling constant).
According to Eq.~(\ref{brdf}), this effect will not change the branching ratios.
However, it may negatively affect the measured fluorescence signal (cf. Eq.~(\ref{sel})).
\begin{figure}[hbt!]
 \centering
 \includegraphics[width=3.33in]{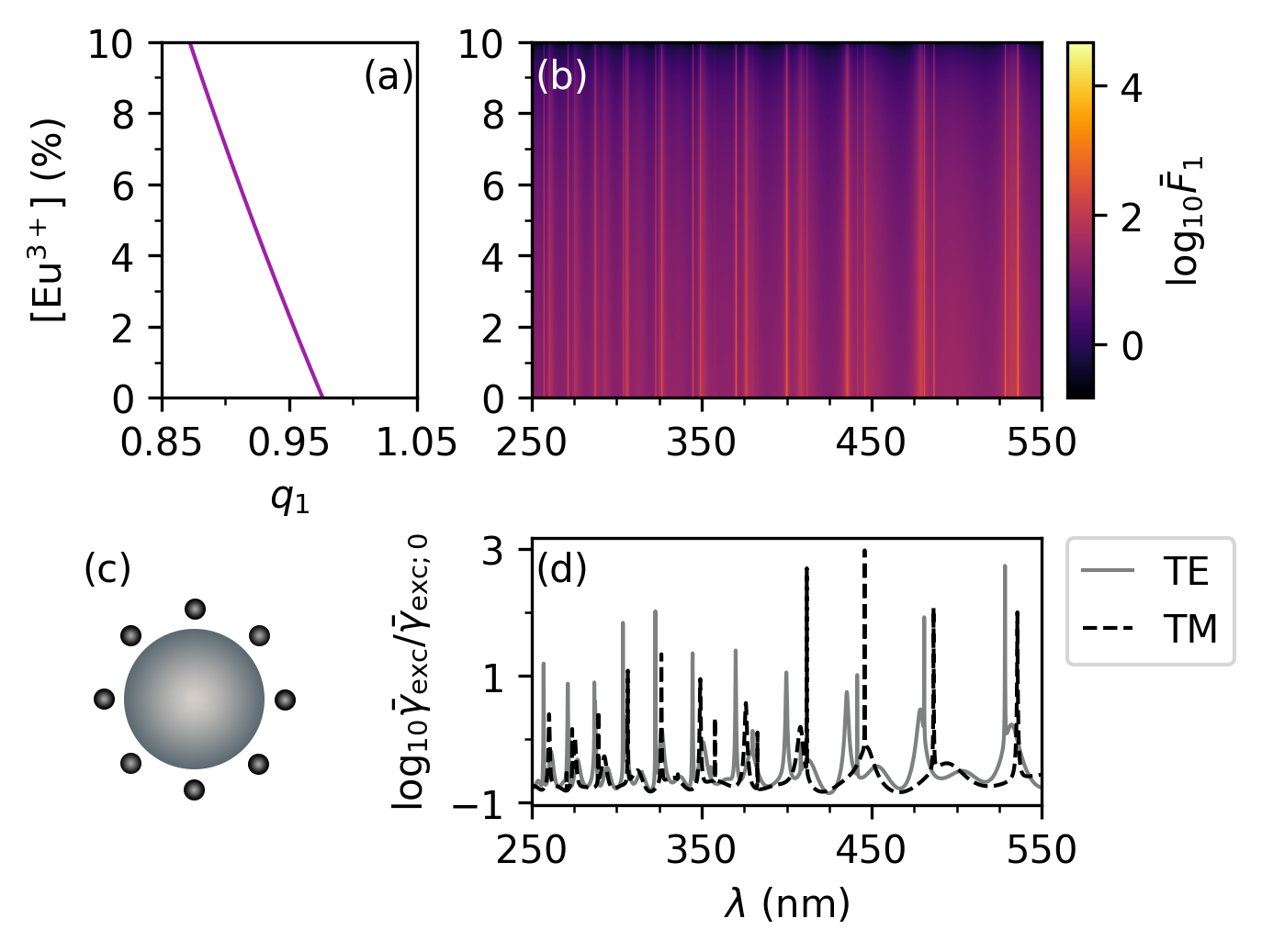}
 \caption{(a) An estimate of expected qualitative behavior of the quantum efficiency $q_1$, according to the Eq.~(\ref{etad}), which depends on the quencher concentration,
 (b) averaged fluorescence enhancement,
 (c) averaged position of emitter and,
 (d) respective averaged excitation enhancement for TE and TM electric modes.
 Radius of the sphere is $r_s=359$~nm and emitters are located at $r_d/r_s=1$.
 Since the emitter is located close to sphere's surface, the local field correction~\cite{Dolgaleva2012} is inapplicable.}
 \label{fig:q_eff_F}
\end{figure}
In order to assess the effect \textit{qualitatively}, we have assumed the same quencher concentration [$Q$] as for silica in Ref.~\cite{DeDood2001}.
Quenchers are typically reactive OH bonds, and [$Q$] is roughly the same for all particles prepared by colloidal synthesis.
The detailed procedure for estimating the coupling constant of Eu$^{3+}$ emitters is presented in Appendix.
The expected qualitative behavior of the concentration quenching is displayed in Fig.~\ref{fig:q_eff_F}(a) with $q_1$ values (see Eq.~(\ref{etad})) depending on Eu$^{3+}$ concentration, which allows us to qualitatively assess the interaction of the emitters.

Finally, it is crucial to gain insight into the possible values of ultimate fluorescence enhancement for the MD transition ($j=1$), which is presented in Fig.~\ref{fig:q_eff_F}(b).
We take the excitation enhancement to be proportional to the surface \textit{averaged} (at fixed radial distance $r_d$) intensity of the electric field, $\bar \gm_{\rm exc}\propto \oint \lvert \vE\rvert^2 \, d\vS$ (see Ref.~\cite{Rasskazov19JOSAA} for the respective closed-form expressions), and intrinsic quantum efficiency of the MD transition is $q_{1;0}\approx0.2$ according to Table~\ref{tab:PL}.
Being substituted in Eq.~(\ref{f}), these values result in the surface \textit{averaged} fluorescence enhancement $\bar F_1$, which can be further increased by placing the emitter in hot spots.
Since the excitation paths strongly depend on the emitter-host configuration (cf. Table~\ref{tab:MD_ions}), a sufficiently broad wavelength range is scanned to obtain a general understanding of the achievable excitation gain, which is shown in Fig.~\ref{fig:q_eff_F}(c).
In short, for lossless particles, exceptional electric field enhancement can be achieved at short wavelengths.
The latter arises from high-order resonances (cf. Figs.~\ref{fig:TiO2rad}(a,b) showing a resonance at $\ld=587$~nm), which together with an insignificant concentration quenching effect can lead to $\bar F$ as large as $\bar F\approx 10^4$, even for a significantly high concentration of Eu$^{3+}$ emitters.
One has also an additional freedom in selecting a suitable excitation wavelength $\ld_{\rm exc}$ which can be employed to optimize results even further.
Clearly, the presence of losses in a particle inevitably suppresses local field enhancement and thus $\gm_{\rm exc}$. 
Nevertheless, the proper choice of MD emitter, particle material, and excitation path can lead to significant enhancement of MD fluorescence after careful engineering of particle size and emitter position.

\section{Conclusions}
Magnetic light fluorescence is fundamentally different from electric light fluorescence:
(i) it cannot be described by a two-level approximation involving only magnetic dipole transitions, and, because magnetic dipole transition from a particular excited level does not occur in isolation,
(ii) one faces a conundrum of resolving the competition of a magnetic transition from a set of different electric dipole transitions, all from a particular excited level (see Figs.~\ref{fig:scheme}(c)-(d)).
The latter requires to calculate for each individual transition its own radiative decay enhancement factor at the corresponding wavelength (see the Jablonski diagram in Fig.~\ref{fig:scheme}(c) and Table~\ref{tab:PL}).
Any enhancement of magnetic light fluorescence is then faced with the difficult task of identifying the regions where electric light fluorescence becomes negligible compared to magnetic light fluorescence.
Despite the above difficult challenges, promising schemes to promote fluorescence of magnetic light due to the MD transitions of trivalent rare-earth ions located inside or near dielectric homogeneous spheres have been discovered. 

Since the excitation enhancement amplifies all transitions by the same factor, the main objective of the sphere is to exert control over the emission processes and influence relevant branching ratios. 
A number of configurations involving the sphere parameters and the rare earth emitter radial position have been identified 
where the branching ratio of the MD transition, $\eta_1$, approaches its limit value of unity under realistic conditions.
The latter means that transitions from a certain initial excited level (e.g., $^5$D$_0$ level of Eu$^{3+}$) are completely dominated by the MD transition.
This is truly remarkable and has never been reported before. Consequently, for properly constructed lossless dielectric spheres and for certain emitter positions, the respective values of the radiation rates and fluorescence gains for MD transitions can be larger than $10^3$ and $10^4$, respectively.
The dimensionless directivity of MD emission can reach the value of $26$.
We strongly believe that the results presented here could serve as an important tool for extending the functionality of photonic devices by promoting magnetic light.

\begin{acknowledgments}
The work by Anton D. Utyushev, Roman Gaponenko and Alexey Shcherbakov (numerical simulations) was supported by the Russian Science Foundation, Grant No. 22-11-00153;
Song Sun would acknowledge support from National Natural Science Foundation of China, Grant No. 62005256 and China Academy of Engineering Physics Innovation and Development Grant No. CX20200011.
\end{acknowledgments}

\begin{widetext}
\appendix
\section{$E\leftrightarrow M$ symmetry in SI units\label{sc:dpemsym} }
For an electric dipole (see Eqs.~(9.4), (9.5) and (9.18) in Ref.~\cite{Jackson1999})
\begin{equation}
 \begin{split}
 \vH_{\rm ed} =& \fr{ck^2}{4\pi} (\vn\times \vp) \, \fr{e^{ikr}}{r} \left(1- \fr{1}{ikr}\right) \ , \\
 \vE_{\rm ed} =& \fr{1}{4\pi\veps} \left\{ k^2 (\vn\times \vp)\times \vn \, \fr{e^{ikr}}{r} \right.  \left.+ \left[3\vn(\vn\cdot\vp)-\vp\right] \left(\fr{1}{r^3}- \fr{ik}{r^2} \right) \right\} \fr{e^{ikr}}{r}\cdot
 \end{split}
\label{dpelmgfsi}
\end{equation}

Given the Maxwell's equations in material medium (see Eqs.~(7.11) and (7.11') in Ref.~\cite{Jackson1999}) (and the vacuum case of Ref.~\cite{Jackson1999}, see Eqs.~(9.4) and~(9.5)),
\begin{equation}
 \begin{split}
 \vH = & \fr{1}{\mu}\, \vnab\times\vA = - \fr{i}{Zk}\,\vnab\times\vE, \\
 \vE = & \fr{iZ}{k}\, \vnab\times\vH ,
 \end{split}
\label{bmeqs}
\end{equation}
after the $\vp\to \vm/c$ substitution, $\vH_{\rm md}$ ($\vE_{\rm md}$) for a magnetic dipole source will be $\vE_{\rm ed}/Z$ ($-Z \vH_{\rm ed}$).
Indeed, for a magnetic dipole (see Eqs.~(9.33) in Ref.~\cite{Jackson1999})
\bg
\vA(\vvr) =\fr{ik\mu}{4\pi}\,
(\vn\times\vm)\, 
\fr{e^{ikr}}{r}\,\left(1- \fr{1}{ikr}\right).
\label{Amd}
\eg
On comparing with the first Eqs.~(\ref{bmeqs}), $\vA$ for the magnetic dipole is thus $(i\mu/k) \vH_{\rm ed}$.
On making use of the latter in the first of Eq.~(\ref{bmeqs}), one finds with the help of  Maxwell's equations in a material medium, 
\begin{equation}
 \begin{split}
 \vH_{\rm md} =& \fr{i\mu}{\mu k}\, \vnab\times \vH_{\rm ed}
 = \fr{i}{k}\fr{k}{iZ}\, \vE_{\rm ed} = \fr{1}{Z}\, \vE_{\rm ed} \ , \\
 \vE_{\rm md} =& \fr{iZ}{k}\, \vnab\times\vH_{\rm md} = \fr{iZ}{k}\, \fr{1}{Z}\, \vnab\times \vE_{\rm ed}  = \fr{i}{k}\, \vnab\times \vE_{\rm ed} = -Z \vH_{\rm ed}.
 \end{split}
\label{emdds}
\end{equation}
The above expressions reflect a \textit{duality} symmetry of the Maxwell's equations in the dipole case. A straightforward consequence is that
\bg
 (\vE_{\rm md} \times \vH_{\rm md}) = (\vE_{\rm ed} \times \vH_{\rm ed}),
\eg
i.e., formally, neither the magnitude nor the orientation of the Poynting vector is changed (up to the substitution $\vp\to \vm/c$). 
Therefore the decay rates (normalized to those of a free dipole) will remain the same as when calculated with $\tl \vE_{\rm d}$ and $\tl \vH_{\rm d}$ by simply interchanging the $E$ and $M$ mode labels ($E\leftrightarrow M$) in the decay rates formulas of an electric dipole case.

There is no other way to interpret a MD than as a circulating current.
Therefore the $\rm {ED}\leftrightarrow \rm {MD}$ duality is a special duality case not covered by the conventional duality of the vacuum and the source Maxwell's equations, the latter assuming the presence of magnetic monopoles.

\section{Radiative and nonradiative decay rates of electric and magnetic dipoles in a presence of a homogeneous sphere\label{sc:mdrates}}
Provided well-established results for radiative and nonradiative decay rates of \textit{electric} dipole emitter located inside or outside a sphere at $r_d$ distance from a center of a sphere (normalized with respect to $\Gm_{\rm rad;ed;0}$, the intrinsic radiative decay rate in the absence of a sphere)~\cite{Moroz2005},

\begin{equation}
\begin{split}
\tl{\Gm}^{\perp}_{\rm rad;ed} = 
\dfrac{\Gm^{\perp}_{\rm rad;ed}}{\Gm_{\rm rad;ed;0}}
& = \dfrac{3}{2 x_d^4} {\cal N}_{\rm rad} \sum_{l=1}^{\infty}l(l+1)(2l+1) \left| {\cal F}_{El}(x_d) \right|^2 \ , \\
\tl{\Gm}^{\parallel}_{\rm rad;ed} = 
\dfrac{\Gm^{\parallel}_{\rm rad;ed}}{\Gm_{\rm rad;ed;0}} 
& = \dfrac{3}{4 x_d^2} {\cal N}_{\rm rad} \sum_{l=1}^{\infty}(2l+1) \left[ \left| {\cal F}_{Ml}(x_d) \right|^2 + \left| {\cal F}^{\;\prime}_{El}(x_d) \right|^2 \right] \ , \\
\tl{\Gm}^{\perp}_{\rm nrad;ed} = 
\dfrac{\Gm^{\perp}_{\rm nrad;ed}}{\Gm_{\rm rad;ed;0}} 
& =   \mb{Im} (\veps_s)\,\dfrac{3k^3_d}{2x_d^4} {\cal N}_{\rm nrad} \sum_{l=1}^{\infty} l(l+1)(2l+1) I_{El} \left| \zeta_l \left(x_d\right) \right|^2 \ , \\
\tl{\Gm}^{\parallel}_{\rm nrad;ed} = 
\dfrac{\Gm^{\parallel}_{\rm nrad;ed}}{\Gm_{\rm rad;ed;0}} 
& =   \mb{Im} (\veps_s)\, \dfrac{3k^3_d}{4x_d^2} {\cal N}_{\rm nrad} \sum_{l=1}^{\infty} (2l+1) \left[ I_{Ml} \left| \zeta_l \left(x_d\right) \right|^2 + I_{El} \left| \zeta_l \left(x_d\right) \right|^2 \right] \ ,
\end{split}
\label{eq:EDdecay}
\end{equation}

one finds straightforwardly upon using $E\leftrightarrow M$ symmetry from Eq.~(\ref{emdds}) the respective expressions for a \textit{magnetic} dipole:

\begin{equation}
\begin{split}
\tl{\Gm}^{\perp}_{\rm rad;md} = 
\dfrac{\Gm^{\perp}_{\rm rad;md}}{\Gm_{\rm rad;md;0}}
& = \dfrac{3}{2 x_d^4} {\cal N}_{\rm rad} \sum_{l=1}^{\infty}l(l+1)(2l+1) \left| {\cal F}_{Ml}(x_d) \right|^2 \ , \\
\tl{\Gm}^{\parallel}_{\rm rad;md} = 
\dfrac{\Gm^{\parallel}_{\rm rad;md}}{\Gm_{\rm rad;md;0}} 
& = \dfrac{3}{4 x_d^2} {\cal N}_{\rm rad} \sum_{l=1}^{\infty}(2l+1) \left[ \left| {\cal F}_{El}(x_d) \right|^2 + \left| {\cal F}^{\;\prime}_{Ml}(x_d) \right|^2 \right] \ , \\
\tl{\Gm}^{\perp}_{\rm nrad;md} = 
\dfrac{\Gm^{\perp}_{\rm nrad;md}}{\Gm_{\rm rad;md;0}} 
& =  \mb{Im} (\veps_s)\, \dfrac{3k^3_d}{2x_d^4} {\cal N}_{\rm nrad} \sum_{l=1}^{\infty} l(l+1)(2l+1) I_{Ml} \left| \zeta_l \left(x_d\right) \right|^2 \ , \\
\tl{\Gm}^{\parallel}_{\rm nrad;md} = 
\dfrac{\Gm^{\parallel}_{\rm nrad;md}}{\Gm_{\rm rad;md;0}} 
& =  \mb{Im} (\veps_s)\, \dfrac{3k^3_d}{4x_d^2} {\cal N}_{\rm nrad} \sum_{l=1}^{\infty} (2l+1) \left[ I_{El} \left| \zeta_l \left(x_d\right) \right|^2 + I_{Ml} \left| \zeta_l \left(x_d\right) \right|^2 \right] \ ,
\end{split}
\label{eq:MDdecay}
\end{equation}
where $x_d = k_d r_d$, $r_d$ is the distance from the center of a sphere to an \textit{electric} (\textit{magnetic}) dipole location, $k_d = 2\pi n_d/\ld$ is the wavevector in the medium where the emitting dipole is located ($n_d=n_s$ in case if $r_d<r_s$ and $n_d=n_h$ if $r_d>r_s$), $l$ denotes the orbital angular moments (in other words, multipole number), $\zeta_l(x) = x h^{(1)}_l(x)$ is the Riccati-Bessel function with $h^{(1)}$ being the spherical Hankel function of the first kind, and prime denotes the derivative with respect to the argument in parentheses.
Expressions for coefficients ${\cal N}_{\rm rad}$ and ${\cal N}_{\rm nrad}$, functions ${\cal F}_{pl}(x_d)$ and ${\cal D}_{pl;a}(x_d)$, and radial integrals $I_{pl}$ are provided below for convenience.
For the respective detailed derivation, we refer the reader to Ref.~\cite{Moroz2005}.
As a reminder, subscript $p=E$ corresponds to electric (TM) mode, and $p=M$ corresponds to magnetic (TE) mode.

The coefficients ${\cal N}_{\rm rad}$ and ${\cal N}_{\rm nrad}$ in Eqs.~(\ref{eq:EDdecay}) and~(\ref{eq:MDdecay}) depend on whether the decay rates were normalized with respect to the radiative decay rates in infinite homogeneous medium having the refractive index of 
(i) the \textit{host} or 
(ii) the medium where the \textit{dipole} is located, whether it is the host or the sphere:

\begin{equation}
\begin{split}
 {\cal N}^{\rm host}_{\rm rad} & = \dfrac{n_d^3}{\veps_d} \dfrac{\veps_h}{n^3_h} \ , \quad 
 {\cal N}^{\rm dip}_{\rm rad} = \left(\dfrac{n_d}{n_h}\right)^6 \left( \dfrac{\veps_h}{\veps_d} \right)^2 \ , \\
 {\cal N}^{\rm host}_{\rm nrad} & = \dfrac{n_d^3}{n^3_h} \dfrac{\veps_h}{\veps_d^2} \ , \quad 
 {\cal N}^{\rm dip}_{\rm nrad} = \dfrac{1}{\veps_d} \cdot
 \end{split}
 \label{eq:dcynorm}
\end{equation}
The functions ${\cal F}_{pl}(x_d)$ and ${\cal D}_{pl;a}(x_d)$ in Eqs.~(\ref{eq:EDdecay}) and~(\ref{eq:MDdecay}) depend on the relative position of the emitter with respect to the sphere (inside or outside). 
In terms of the Riccati-Bessel functions $\psi_l$ and $\zeta_l$,
\begin{equation}
 {\cal F}_{pl}(x_d) = 
 \begin{cases}
 \dfrac{\psi_l(x_d)}{T_{21;pl}^-} \ , & {\rm inside} \ , \\[10pt]
 \psi_l (x_d)+\dfrac{T_{21;pl}^+}{T_{11;pl}^+} \, \zeta_l(x_d), 
 & {\rm outside}.
 \end{cases}
\end{equation}
where $\psi_l(x) = xj_l(x)$ is the Riccati-Bessel function with $j_l(x)$ being the spherical Bessel function of the first kind, and the numeric subscript, e.g. ``21'', of the transfer-matrix $T$ (given below) corresponds to its respective element.

The respective radial integrals $I_{pl}$ in Eqs.~(\ref{eq:EDdecay}) and~(\ref{eq:MDdecay}) are
\begin{equation}
 \begin{split}
 I_{Ml} = & \dfrac{1}{|k_s|^2} \int \left| {\cal A}_{Ml} \psi_l(k_s r)\right|^2 {\rm d}r \ , \\
 I_{El} = & \dfrac{l(l+1)}{|k_s|^4} \int \left| {\cal A}_{El} \psi_l(k_s r)\right|^2 \dfrac{{\rm d}r}{r^2}+\dfrac{1}{|k_s|^2} \int_{r_s} \left| {\cal A}_{El} \psi^{\prime}_{l}(k_s r)\right|^2 {\rm d}r \ .
 \end{split}
 \label{eq:Iabs}
\end{equation}
Here $k_s = 2\pi n_s/\ld$ is the wave vector in the sphere medium, and coefficients ${\cal A}_{pl}$ are
\begin{equation}
 {\cal A}_{pl} = 
 T_{11;pl}^- + T_{12;pl}^- \dfrac{T_{21;pl}^+}{T_{11;pl}^+} \cdot
 \label{eq:AIabs}
\end{equation}

Finally, the respective backward and forward transfer matrices for the electric and magnetic modes:

\begin{equation}
T^-_{Ml} =
- i \begin{pmatrix}
  \tilde{n} \zeta_l'(x_s) \psi_l(\tilde{x}_s) - \tilde{\mu} \zeta_l(x_s)\psi_l'(\tilde{x}_s) 
& \tilde{n} \zeta_l'(x_s) \zeta_l(\tilde{x}_s) - \tilde{\mu} \zeta_l(x_s) \zeta_l'(\tilde{x}_s) \\
- \tilde{n} \psi_l'(x_s)\psi_l(\tilde{x}_s) + \tilde{\mu} \psi_l(x_s)\psi_l'(\tilde{x}_s)
& - \tilde{n} \psi_l'(x_s) \zeta_l(\tilde{x}_s) + \tilde{\mu} \psi_l(x_s)\zeta_l'(\tilde{x}_s)
\end{pmatrix} \ ,
\label{eq:bacmtm}
\end{equation}

\begin{equation}
T^-_{El} = 
- i \begin{pmatrix}
  \tilde{\mu} \zeta_l'(x_s)\psi_l(\tilde{x}_s) - \tilde{n} \zeta_l(x_s)\psi_l'(\tilde{x}_s) 
& \tilde{\mu} \zeta_l'(x_s) \zeta_l(\tilde{x}_s) - \tilde{n} \zeta_l(x_s) \zeta_l'(\tilde{x}_s) \\
- \tilde{\mu} \psi_l'(x_s)\psi_l(\tilde{x}_s) + \tilde{n} \psi_l(x_s)\psi_l'(\tilde{x}_s)
& - \tilde{\mu} \psi_l'(x_s) \zeta_l(\tilde{x}_s) + \tilde{n} \psi_l(x_s)\zeta_l'(\tilde{x}_s)
\end{pmatrix} \ ,
\label{eq:bacetm}
\end{equation}

\begin{equation}
T^+_{Ml} = 
- i \begin{pmatrix}
  \zeta_l'(\tilde{x}_s) \psi_l(x_s)/\tilde{n} - \zeta_l(\tilde{x}_s)\psi_l'(x_s)/\tilde{\mu}
& \zeta_l'(\tilde{x}_s) \zeta_l(x_s)/\tilde{n} - \zeta_l(\tilde{x}_s)\zeta_l'(x_s)/\tilde{\mu} \\
- \psi_l'(\tilde{x}_s)\psi_l(x_s)/\tilde{n} + \psi_l(\tilde{x}_s)\psi_l'(x_s) /\tilde{\mu}
& - \psi_l'(\tilde{x}_s) \zeta_l(x_s)/\tilde{n} + \psi_l(\tilde{x}_s)\zeta_l'(x_s)/\tilde{\mu}
\end{pmatrix} \ ,
\label{eq:formtm}
\end{equation}

\begin{equation}
T^+_{El} =
- i \begin{pmatrix}
  \zeta_l'(\tilde{x}_s)\psi_l(x_s)/\tilde{\mu}  - \zeta_l(\tilde{x}_s)\psi_l'(x_s) /\tilde{n}
& \zeta_l'(\tilde{x}_s) \zeta_l(x_s)/\tilde{\mu} - \zeta_l(\tilde{x}_s) \zeta_l'(x_s) /\tilde{n} \\
- \psi_l'(\tilde{x}_s)\psi_l(x_s)/\tilde{\mu} + \psi_l(\tilde{x}_s)\psi_l'(x_s) /\tilde{n}
& - \psi_l'(\tilde{x}_s) \zeta_l(x_s)/\tilde{\mu} + \psi_l(\tilde{x}_s)\zeta_l'(x_s) /\tilde{n}
\end{pmatrix} \ ,
\label{eq:foretm}
\end{equation}
where prime again denotes the derivative with respect to the argument in parentheses, $x_s=k_s r_s$ and $\tilde{x}_s=x_s / \tilde{n}=k_h r_s$ are the internal and external dimensionless size parameters, and $\tilde{n}=n_s/n_h$ and $\tilde{\mu}=\mu_s/\mu_h$ are sphere's relative refractive index and permeability normalized to that of the host.

\section{Fluorescence quantum efficiency in the presence of nonradiative losses\label{sseceq7}}
Equation~(\ref{etad}) for the fluorescence quantum efficiency of the main text,
\bg
q_j = \fr{\Gm_{{\rm rad};j}}{\Gm_{\rm tot}},
\nn   
\eg
contains \textit{absolute} decay rates.
Provided there is some nonradiative decay rate $\Gm_{{\rm nrad}}$ involved, $\Gm_{\rm tot}$ in Eq.~(\ref{etad}) is
\bg
\Gm_{\rm tot}=\Gm_{{\rm nrad}} + \sum_j  \Gm_{{\rm rad};j}\cdot
\eg
In order to make use of Eqs.~(\ref{eq:EDdecay}) and~(\ref{eq:MDdecay}), Eq.~(\ref{etad}) has to be recast in a more convenient form. 
Divide both the numerator and denominator of Eq.~(\ref{etad}) with some arbitrary predetermined radiative rate $\Gm_R$, 
\bg
q_j = \fr{\Gm_{{\rm rad};j}/\Gm_R}{\Gm_{\rm tot}/\Gm_R}\cdot
\label{etadm}
\eg
With the knowledge of that the time-averaged total radiated power of a free dipole of dipole moment $\vp$ is~\cite{Jackson1999}
\bg
P_{tot}=\fr{ck_h^4 |\vp |^2 }{3 \veps_h n_h},
\eg
$\Gm_R$ can be taken as the radiative decay rate of a free dipole of a unit dipole moment,
\bg
\Gm_R=\fr{P_{tot}}{\hbar\om}=\fr{k_h^3}{3\hbar\veps_h},
\eg
where we have used that $k_h/\om=n_h/c$.
In the expressions above, $k_h=2\pi/\ld_h$, $\veps_h$, and $n_h$ are the wave vector, the dielectric constant, and refractive index in the host medium. 
In principle, one can use for $\Gm_R$ any fixed decay rate.
With a single radiative channel, one can choose $\Gm_R=\Gm_{{\rm rad}}^0$, whereby one recovers the usual expressions.

Obviously, one can recast the ratio
$\Gm_{\rm tot}/\Gm_R$ as
\bg
\Gm_{\rm tot}/\Gm_R=\Gm_{{\rm nrad}}/\Gm_R + 
\sum_k  (\Gm_{{\rm rad};k}/\Gm_{{\rm rad};k}^0)\cdot (\Gm_{{\rm rad};k}^0/\Gm_R).
\eg
The first parenthesis in the sum is provided by Eqs.~(\ref{eq:EDdecay}) and~(\ref{eq:MDdecay}). 
The second parenthesis in the sum can be determined from Table~\ref{tab:PL}.
With the numerator of Eq.~(\ref{etadm}) recast analogously as
$(\Gm_{{\rm rad};j}/\Gm_{{\rm rad};j}^0)\cdot (\Gm_{{\rm rad};j}^0/\Gm_R)$, such a modified Eq.~(\ref{etadm}) has been used in our calculations.

\section{Coupling constant for multiple Eu$^{3+}$ emitters}
Coupling constant, $C_{\text{Eu-Eu}}$, entering the expression for nonradiative decay rates due to concentration quenching,
\begin{equation}
    {\Gm}_{\rm nrad} = 8\pi C_{\rm Eu-Eu}[{\rm Eu}^{3+}][Q] \ ,
\end{equation} 
can be estimated from the experimental measurements for $^5$D$_0$ level lifetime as a function of Eu$^{3+}$ doping concentration (see Fig. 4 in Ref.~\cite{Dordevic2013}).
Excited state lifetime is known to be inversely proportional to the total decay rate: 
\begin{equation}
    \tau \propto \frac{1}{\Gm_{\rm tot}} \propto \frac{1}{\Gm_R + \Gm_{\rm nrad} } \cdot
\end{equation}
With known $\Gm_R = 891{\rm s}^{-1}$~\cite{Lima2016} and fixed quencher concentration [$Q$], one can approximate experimental data for $\tau$ with a linear fit.
For the data presented in Ref.~\cite{Dordevic2013}, we have found $ 8\pi C_{\rm Eu-Eu}[Q]\approx 250\frac{1}{{\rm at.}\% \: {\rm s}}$, which is the value used in our simulations.

\section{Directivity of a magnetic dipole in the presence of a sphere\label{sc:directivity}}
By definition, the directivity $\mathcal{D}$ is a relation of the power emitted into a certain direction to the solid angle averaged emitted power:
\bg
\DD=\fr{4\pi r^2 |{\bf E} ({\bf r})|^2}{r^2 \oint |{\bf E} ({\bf r})|^2 d\Om}\cdot
\label{Ddf}
\eg

In the presence of a spherical particle, the magnetic dipole radiation pattern is modified.
The respective electric and magnetic fields can be found upon using $E\leftrightarrow M$ symmetry:
\begin{eqnarray}
{\bf H}_{\rm md} ({\bf r}) & = & \sqrt{\dfrac{\varepsilon}{\mu}} \sum_L \left[
 D_{ML}{\bf H}_{ML}(k,{\bf r})+ D_{EL}{\bf H}_{EL}(k,{\bf r}) \right],
\\
{\bf E}_{\rm md} ({\bf r}) & = &i\, \sum_L \left[ 
D_{ML} {\bf H}_{EL}(k,{\bf r})+ D_{EL} {\bf H}_{ML}(k,{\bf r})\right],
\label{dipemf}
\end{eqnarray}
Where the amplitudes $D_{pL}$ of the outgoing radiated field are given as follows, depending on the location of the dipole with respect to a sphere:

\begin{equation}
 D_{p L} = 
\begin{cases}
\dfrac{a^>_{p L}}{T^-_{22;p l} } \ , & {\rm inside} \ , \\[10pt]
a^>_{p L} + \dfrac{T^+_{21;p l}}{T^+_{11;p l}} a^<_{p L} \ , & {\rm outside} \ , \nn
\end{cases}
\end{equation}

with

\begin{eqnarray}
a^<_{ML} &=& 
4\pi i (k^3/c\varepsilon) \, {\bf m}\cdot {\bf H}_{ML}^*(k,{\bf r}_d),
\nonumber\\
a^<_{EL} &=& 4\pi i (k^3/cn) \, \sqrt{\frac{\mu}{\varepsilon}}
\, {\bf m}\cdot {\bf H}_{EL}^*(k,{\bf r}_d)
\nonumber\\
 &=& 4\pi i (k^3/c\varepsilon) \, {\bf m}\cdot{\bf H}_{EL}^*(k,{\bf r}_d).
\label{alphas}
\end{eqnarray} 

and

\begin{eqnarray}
a^>_{ML} &=& 
4\pi i (k^3/c\varepsilon) \, {\bf m}\cdot {\bf J}_{ML}^*(k,{\bf r}_d),\nonumber\\
a^>_{EL} &=& 4\pi i (k^3/cn) \, \sqrt{\frac{\mu}{\varepsilon}}
 \, {\bf m}\cdot {\bf J}_{EL}^*(k,{\bf r}_d)
\nonumber\\
 &=& 4\pi i (k^3/c\varepsilon) \, {\bf m}\cdot {\bf J}_{EL}^*(k,{\bf r}_d).
\label{as}
\end{eqnarray}
The star here and above denotes the complex conjugation, which only applies to the vector spherical harmonics and not to the spherical Bessel functions~\cite{Chew1987,Chew1988}.

Normalized transverse vector multipoles are defined as
\begin{eqnarray}
{\bf F}_{ML}(k_n,{\bf r}) &= & f_{ML} (k_n r) {\bf Y}^{(m)}_L({\bf r}),
\nonumber\\
{\bf F}_{EL}(k_n,{\bf r}) &=&
\frac{1}{k_n r}\left\{\sqrt{l(l+1)}f_{EL} (k_n r){\bf Y}^{(o)}_L({\bf r}) 
\right.
\nn\\
&& \left. 
+
\dfrac{{\rm d}}{{\rm d}r}\left(r f_{EL}(k_n r)\right)\,{\bf Y}^{(e)}_L({\bf r})\right\},
\label{fvmultip}
\end{eqnarray}
where $f_{\gamma L}$ is an arbitrary linear combination of spherical Bessel functions (${\bf F}_{\gm L} \equiv {\bf J}_{\gm L}$ for $f_{\gm L} = j_l$ and ${\bf F}_{\gm L} \equiv {\bf H}_{\gm L}$ for $f_{\gm L} = h^{(1)}_l$), ${\bf Y}^{(a)}_L$ are vector spherical harmonics, and $L$ is a composite angular momentum index, $L=(l,m)$, where the respective $l\ge 1$ and $m$ label the orbital and magnetic angular numbers. 

In what follows, it is expedient to introduce scalar angular functions,
\begin{eqnarray}
\pi_{ml}(\theta)=\frac{m}{\sin\theta}\, d_{0m}^l(\theta), \quad
\tau_{ml}(\theta)=\frac{\rm d}{{\rm d}\theta}\, d_{0m}^l (\theta),
\label{pitau}
\end{eqnarray}
which are defined in terms of the Wigner $d$-functions $d_{0m}^l$ and
which can all be generated by stable recurrences. Then with the orthonormal spherical coordinate basis vectors $\ve_r, \ve_\theta, \ve_\phi$:
\bea
\vY^{(m)}_L &=& (-1)^m i d_l \left(i \ve_\theta \,\pi_{ml} - \ve_\phi  \,
\tau_{ml} \right) e^{im\vphi}
\nn\\
&=& 
i (-1)^m d_l {\bf C}_L(\theta) e^{im\varphi},
\nn\\
\vY^{(e)}_L &=& (-1)^m i d_l \left( \ve_\theta\,\tau_{ml} + i \ve_\phi \,
\pi_{ml} \right) e^{im\vphi}
\nn\\
&=&
i (-1)^m d_l {\bf B}_L(\theta) e^{im\varphi} ,
\nn\\
\vY^{(o)}_L &=& i \ve_r Y_L=i \gm_L'\, P_l^m (\cos\theta)\, e^{im\vphi} \ve_r
\nn\\
&=& (-1)^m i \sqrt{l(l+1)}\, d_l \, d_{0m}^l (\theta)\, e^{im\vphi} \ve_r
\nn\\
&=& (-1)^m i \sqrt{l(l+1)}\, d_l \, {\bf P}_L (\theta)\, e^{im\vphi},
\label{vectsphec}
\eea
where $Y_L$ with $l\ge 1$ are the usual orthonormal scalar spherical harmonics in the Condon-Shortley convention defined in terms of the associate Legendre functions $P_l^m (\cos\theta)$ as, for instance, by Jackson~\cite{Jackson1999}, and the numerical constant $\gamma_L'$ of Ref.~\cite{Tsang1985} and numerical constant $d_l$ of Ref.~\cite{Mishchenko1991}, are
\begin{equation}
d_l=\left[\frac{2l+1}{4\pi l(l+1)}\right]^{1/2},
\quad
\gamma_L' = \sqrt{ \frac{(2l+1)(l-m)!}{4\pi(l+m)!} }\cdot
\label{dldf}
\end{equation}

$\vC$, $\vB$, and $\vP$ can be expressed as~\cite{Mishchenko1991} 
\bea
\vC_L(\theta) &=&
\ve_\theta \fr{im}{\sin\theta}
d_{0m}^l (\theta) - \ve_\phi \frac{{\rm d} d_{0m}^l (\theta)}{{\rm d}\theta},
\nn\\
\vB_L (\theta) &=&
\ve_\theta \frac{{\rm d} d_{0m}^l (\theta)}{{\rm d}\theta} +\ve_\phi \fr{im}{\sin\theta}
d_{0m}^l (\theta) ={\bf r}_0\times\vC_L,
\nn\\
 \vP_L(\theta) &=& \fr{{\bf r}}{r}\, d_{0m}^l (\theta).
 \label{mishbcp}
\eea
For the special case $\bt=\theta=0$ and $m'=0$
\bg
d_{0m}^l(0) = (-1)^{m} \dt_{0m} = \dt_{m0}.
\label{ed4.2.3}
\eg
It should be remembered though that Edmonds $d_{0m}^l$ are related to our up to prefactor of $(-1)^m$.

In the case of a magnetic dipole near a spherical particle, when both the dipole and the sphere center are located on axis $z$, and when $\theta_0=0$, the general Eq.~(\ref{Ddf}) reduces to:
\begin{equation}
\mathcal{D}_{\rm md} = \frac{\left| \sum_{l} (-i)^l \sqrt{(2l+1)} \left( D_{Ml} + D_{El} \right) \right|^2}{\sum_{l}\left(|D_{Ml}|^2 + |D_{El}|^2\right)} \cdot
\end{equation}
\end{widetext}

\providecommand{\noopsort}[1]{}\providecommand{\singleletter}[1]{#1}%

\end{document}